\documentclass[11pt,a4paper]{article}
\pdfoutput=1
\usepackage{jcappub}
\usepackage{lmodern}
\usepackage{graphicx}
\usepackage{amsmath}
\usepackage{subfig}

\usepackage{cellspace}
\newcommand{\lyxmathsym}[1]{\ifmmode\begingroup\def\b@ld{bold}
  \text{\ifx\math@version\b@ld\bfseries\fi#1}\endgroup\else#1\fi}
\newcommand{\degree}{\ensuremath{^\circ}}

\begin{document}

\title{Leakage of power from dipole to higher multipoles due to non-symmetric beam shape of the CMB missions}

\author{Santanu Das,}
\author{Tarun Souradeep}

\emailAdd{santanud@iucaa.ernet.in}
\emailAdd{tarun@iucaa.ernet.in}

\affiliation{IUCAA, P. O. Bag 4, Ganeshkhind, Pune 411007, India}

\abstract
{A number of studies of WMAP and Planck claimed the low multipole (specially quadrupole) power deficiency in CMB power spectrum. Anomaly in the orientations of the low multipoles have also been claimed. There is a possibility that the power deficiency at low multipoles may not be of primordial origin and is only an observation artifact coming from the scan procedure adapted in the WMAP or Planck satellites. Therefore, it is always important to investigate all the observational artifacts that can mimic them. The CMB dipole which is much higher than the quadrupole can leak to the higher multipoles due to the non-symmetric beam shape of the WMAP or Planck. We observe that a non-negligible amount of power from the dipole can get transferred to the quadrupole and the higher multipoles due to the non-symmetric beam shapes and contaminate the observed measurements. The orientation of the quadrupole generated by this power transfer is surprisingly very close to the quadrupole observed from the WMAP and Planck maps. However, our analysis shows that the orientation of the quadrupole 
can not be explained using only the dipole power leakage. In this paper we calculate the amount of quadrupole power leakage for different WMAP bands. For Planck we present the results in terms of upper limits on asymmetric beam parameters that can lead to significant amount of power leakage.}

\maketitle

\section{Introduction}

The theoretical predictions of the cosmic microwave background (CMB) can provide a very accurate match to the observational results, making the standard model of cosmology a remarkable scientific success. Several ground based and space based experiments are carried out and the precision in the measurements of the CMB has improved dramatically in the past decade. After WMAP and Planck results the observational CMB physics comes to a state where each and every small departure from theoretical model may contribute to new physics. It is observed that the power at the low multipoles of the CMB power spectrum is lower than the theoretical predictions of the best fit model. The measured quadrupole power ($3C_2/\pi$) from WMAP-9 year data is $150.6398\mu K^2$~\cite{WMAPQuad} and Planck data is $299.495 \mu K^2$~\cite{PlanckQuad} where as the theoretical $\Lambda$CDM best fit power for quadrupole is $\sim1221.3 \mu K^2$~\cite{WMAPBFit}. Possible angular correlation between the orientations of the low multipoles has also been claimed \cite{Ade2013,deOliveiraCosta2003,Tegmark2003,Huterer2006,Copi2005}. Although there  can be some cosmological ramification of these anomalies \cite{Feng2003,Jain2009,Das:2013sca,Das:2013gta}, it is important to check all the observational artifacts  that can mimic these. Recently  Moss, Scott and Sigurdson show that the power from dipole can leak to quadrupole due to the pointing offsets of the WMAP beam \cite{Moss-2010}. In this paper we show that a similar effect of leakage of CMB dipole into low CMB multipoles can also arise due to non-symmetric beam shape of the CMB experiments and the power transfer can be as much as $\sim 65.5\mu K^2$ for some frequency band of WMAP.

An accurate measurement of the CMB power spectrum is the primary goal of all the CMB experiments. It needs an accurate data analysis technique. In the WMAP experiment the beam shape of the detectors are not reflectionally symmetric. However, in most of the data analysis techniques true shape of the beam is not taken into account. The CMB dipole, which is much stronger than the quadrupole, may leak to the quadrupole or the higher multipoles due to the effect of non-symmetric beam. However, assuming a circular beam in the data analysis technique, this leakage of power can not be accounted for. Therefore, the contribution of this effect contaminates the resultant map, generated by this inadequate data analysis technique. The effects of non-symmetric beam are discussed by different authors \cite{Fosalba-2002,Das2014a,Joshi2012,Das2015,Mitra-2006,Souradeep-2006,Souradeep-2001,Hinshaw-2007}. However, this particular effect has not been studied in any previous literature.

In this paper, analytical methods are developed to calculate the amount of power leakage and simulations with the actual scan pattern of WMAP are carried out showing the order of power leakage for different WMAP beams. The analysis shows that a non-negligible amount of power can leak from the dipole to quadrupole and the shape of the quadrupole originated by this power transfer is similar to the quadrupole observed by WMAP satellite. However, for WMAP scan strategy the power leakage to octupole or higher multipoles are very less in comparison to the power leaked to quadrupole. A similar analysis is also carried out with Planck scan pattern. We show that for Planck scan pattern the power transfer is not only restricted to quadrupole, but it can also contaminate other higher low multipoles. An upper limit  is set on some beam parameters, that can cause significant power leakage from dipole. Even though the amount of the power leakage is lesser for Planck scan pattern, if not accounted for properly, it may lead to wrong data interpretation. 

The paper is organized as follows. In section~\ref{section2} we provide an analytical description of the beam convolution with a dipole map. In section~\ref{sec3} we briefly present the WMAP scan strategy, beam geometry and the map making procedure from the scanned sky. The results  of our analysis for the dipole power leakage in WMAP is described in section~\ref{sec:WMAP}. Section~\ref{sec:Planck} shows the results of similar analysis for Planck scan. The discussion and conclusion  of the paper is presented in the final section. 

\section{An analytical description of the beam convolution}
\label{section2}

The measured sky temperature in a CMB experiment is a convolution of the real sky temperature with the instrumental beam. If measured temperature along $\gamma_{m}$ direction is $\tilde{T}(\gamma_{m})$, where as the real sky temperature along the direction $\gamma$ is $T(\gamma)$, then they can be related as 

\begin{equation}
\tilde{T}(\gamma_{m})=\int B(\gamma_{m},\gamma)T(\gamma)d\Omega_{\gamma}+T_{n}(\gamma_{m})\;,
\end{equation}

\noindent where, $T_{n}$ is the instrumental noise during the scan procedure. Since we deal with low multipoles,  the noise is ignored in our analysis.  Here, the beam function $B(\gamma_{m},\gamma)$ represents the sensitivity of the telescope around the pointing direction $\gamma_{m}$. This is a two point function and can be expanded in terms of spherical
harmonics as 

\begin{equation}
B(\gamma_{m},\gamma)=\sum_{l=0}^{\infty}\sum_{m=-l}^{l}b_{lm}(\gamma_{m})Y_{l}^{m}(\gamma)\;,
\end{equation}

\noindent where, $Y_{l}^{m}(\gamma)$ are the spherical harmonics and $b_{lm}(\gamma_{m})$ are their coefficients. Since we intend to measure the power leakage from dipole to the quadrupole due to non-symmetric beam, it is convenient to scan a dipole only sky map using a non-symmetric beam and check the amount of power transfer from dipole to the quadrupole and other higher low multipoles in the resultant map. This can give us an estimate of the amount of power leakage from dipole. 

For a dipole only sky map, $T(\gamma)$ can be expressed as a sum of
all the spherical harmonics with $l=1$, i.e. $T(\gamma)=\sum_{m=-1}^{1}a_{1m}Y_{1}^{m}(\gamma)$.
It is always possible to choose a coordinate system such that $a_{1,1}$
and $a_{1,-1}$ modes vanish and the sky temperature can be expressed
as $T(\gamma)=T_{0}Y_{1}^{0}(\gamma)$, where $T_{0}=a_{1,0}$ is
a constant. Then the measured sky temperature along $\gamma_{m}$ can be expressed as 

\begin{eqnarray}
\tilde{T}(\gamma_{m})  =  \int B(\gamma_{m},\gamma)T(\gamma)d\Omega_{\gamma}
  =  T_{0}\sum_{l=0}^{\infty}\sum_{m=-l}^{l}b_{lm}(\gamma_{m})\int Y_{l}^{m}(\gamma)Y_{1}^{0}(\gamma)d\Omega_{\gamma} 
  =  T_{0}b_{10}(\gamma_{m})\;,
\label{eq:scanned temperature}
\end{eqnarray}

\noindent where $\gamma_{m}$ is the direction along which the beam
is oriented. It is always convenient to orient the beam 
along a fixed direction in the sky, say along the $\hat{z}$ direction in 'xyz' coordinate system
and consider the multipole $b_{lm}$ to characterize the beam. In this case we need to calculate the beam spherical
harmonic coefficients only at one direction and then rotate them 
to a particular direction using the Wigner-D functions
\begin{eqnarray}
b_{10}(\gamma_{m}) & = & \sum_{m'=-1}^{1}b_{1m'}(z)D_{0m'}^{1}(\phi_{m},\theta_{m},\rho_{m})\nonumber \\
 & = & b_{1,-1}(\hat{z})D_{0,-1}^{1}(\phi_{m},\theta_{m},\rho_{m})+b_{1,0}(\hat{z})D_{0,0}^{1}(\phi_{m},\theta_{m},\rho_{m})+b_{1,1}  
 (\hat{z})D_{0,1}^{1}(\phi_{m},\theta_{m},\rho_{m})\nonumber \\
 & = & b_{1,-1}(\hat{z})d_{0,-1}^{1}(\theta_{m})e^{i\rho_{m}}+b_{1,0}(\hat{z})d_{0,0}^{1}(\theta_{m})+b_{1,1}(\hat{z})d_{0,1}^{1}(\theta_{m})e^{-i\rho_{m}}\;.\label{eq:beam rotation}
\end{eqnarray}

\noindent Here $D_{mm'}^{l}(\phi_{m},\theta_{m},\rho_{m}) = e^{-im\phi_m}d_{mm'}^{l}(\theta_m)e^{-im'\rho_m}$ is the Wigner D-Matrix and $d_{mm'}^{l}(\theta_m)$ is the Wigner small d-Matrix. ($\phi_{m},\theta_{m},\rho_{m}$)  are the Eular rotation angles of the direction $\gamma_m$, i.e. $\phi_m$ is the rotation in the 'xy' plane, $\theta_m$ is the inclination from the $\hat{z}$ direction and $\rho_m$ is the rotation of the beam~\cite{Souradeep-2001}. As in our expressions $m=0$,  $b_{10}(\gamma_{m})$ is independent on $\phi_m$. Also we know

\begin{equation}
d_{0,0}^{1}(\theta_{m})=\cos\theta_{m}
\;\;\;\;\;\;\;\;\;\;\;
d_{0,1}^{1}(\theta_{m})=-d_{0,-1}^{1}(\theta_{m})=\frac{1}{\sqrt{2}}\sin\theta_{m}\;.\label{eq:d101}
\end{equation}

\noindent Therefore, using the Eq.\,\ref{eq:scanned temperature} and Eq.\,\ref{eq:beam rotation} and the trigonometric
expressions for Wigner small-d functions, the scanned temperature can be expressed as 

\begin{equation}
\tilde{T}(\gamma_{m})=T_{0}b_{10}(\hat{z})\cos\theta_{m}+\sqrt{2}T_{0}\sin\theta_{m}\left[b_{r}(\hat{z})\cos\rho_{m}+b_{i}(\hat{z})\sin\rho_{m}\right]\;.\label{eq:temp}
\end{equation}

\noindent In Eq.\,\ref{eq:beam rotation}, $b_{1,1}$ and $b_{1,-1}$ are complex quantities. But, as the beam is real, the beam spherical harmonic coefficients should satisfy $b_{1,1}^{*}=-b_{1,-1}$. Hence, for simplifying the expressions we use $b_{1,1}=b_{r}+ib_{i}$, i.e. the real and the imaginary parts of $b_{1,1}$, which led us to the Eq.\,\ref{eq:temp}. 

Using Eq.\,\ref{eq:temp} we can calculate the power leakage from dipole to the quadrupole or the higher multipoles. Eq.\,\ref{eq:temp} shows that the first term that is the term with $b_{10}(\hat{z})$ can not contribute to any power leakage from dipole
as it depends only on $\theta_{m}$  and its expression is same as that of the original dipole. The dipole to quadrupole power transfer is only caused by the terms multiplied with $b_{r}(\hat{z})$ or $b_{i}(\hat{z})$. Therefore, if a beam is designed in such a way 
that the $b_{r}(\hat{z})$ or $b_{i}(\hat{z})$ components of the beam are negligibly small then there can not be any dipole to higher multipole power transfers.

\section{An analytical description of WMAP Scan}
\label{sec3}
\subsection{WMAP Scan geometry}

For knowing the amount of power leakage we need to have the details of the scan
geometry from the WMAP satellite and some mathematical tools to calculate
that. The WMAP satellite
follows a well defined scan pattern, in which the pixels near the two
poles are scanned for large number of times from different directions,
whereas those which are near the equator are scanned for lesser
 number of times. WMAP satellite has a pair of telescopic horns for each frequency
band, both of which are about $~70.5\degree$ 
off the symmetry axis. The satellite has a fast spin about this symmetric
axis with the spin period of around $2.2$ minutes. Along with this
fast spin, the spacecraft has a slow precession, $22.5\degree$ 
about the Sun-WMAP line. This precession period is about $1$ hour.
Finally, the satellite follows the rotation of the earth-sun vector and 
it rotates $360\degree$  per year. 

\begin{figure}
\centering
\includegraphics[scale=0.30,trim = 35 0 20 0, clip]{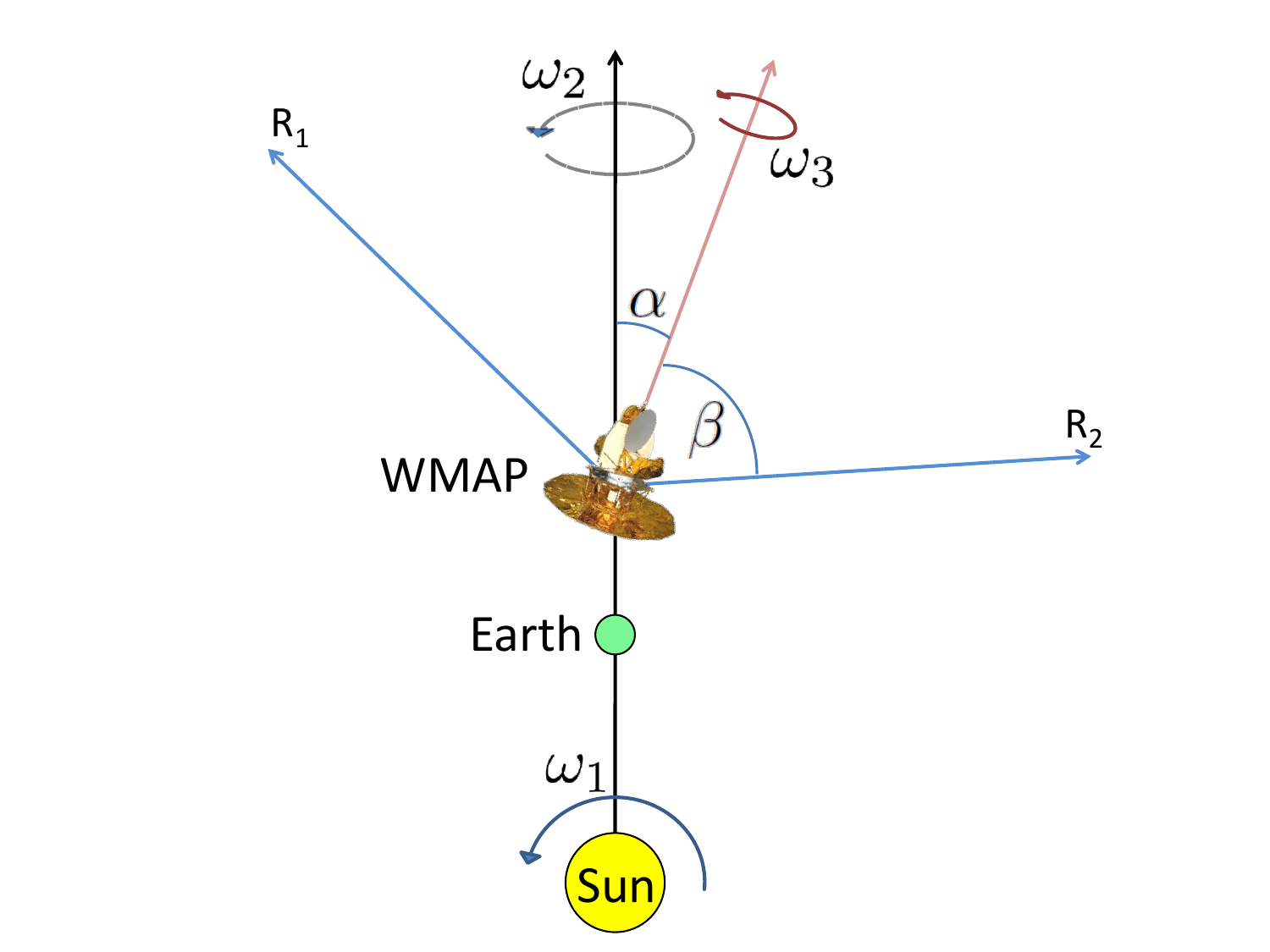}
\includegraphics[scale=0.30,trim = 10 0 100 0, clip]{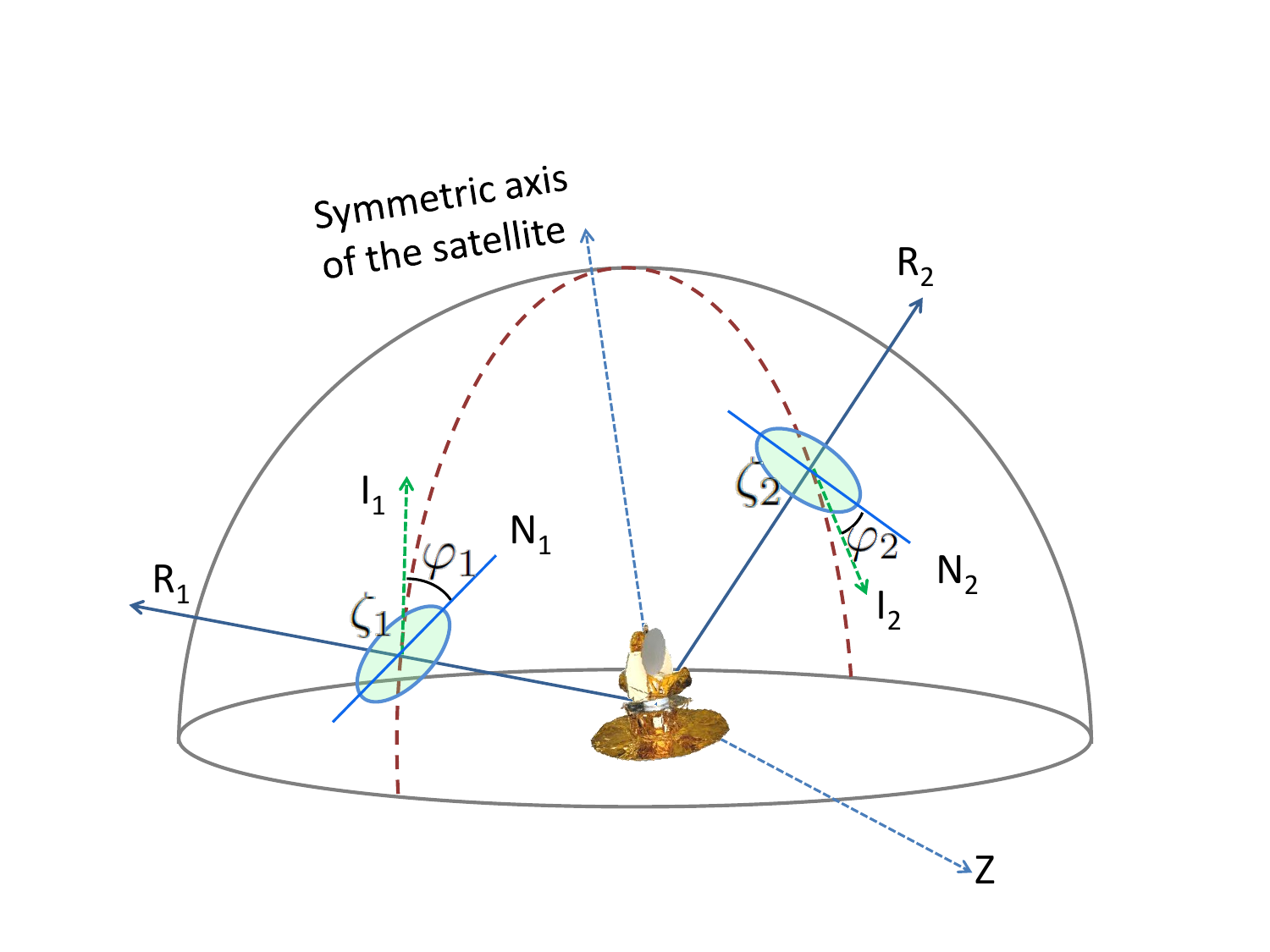}

\caption{\label{fig:WMAPscan-1}Left: WMAP satellite spin and scan strategy is described in the above figure. The blue arrows are
the lines of sight of two horns, red arrow is space craft symmetric axis and
black arrow is the sun-WMAP line.
Right : Beam of the WMAP satellite. $R_{1}$ and
$R_{2}$ are the line of sight of the two beams. $\zeta_{1}$ and $\zeta_{2}$
planes (shown in light blue) are perpendicular to $R_{1}$ and $R_{2}$ respectively. The
beams are located in the $\zeta_1$ and $\zeta_2$ plane; and $N_{1}$
and $N_{2}$ are the direction of the scan axis of the beams.
$\varphi_{1}$ and $\varphi_{2}$ are showing the orientation of $N_{1}$
and $N_{2}$ with respect to $I_{1}$ and $I_{2}$.}
\end{figure}

A schematic diagram describing the scan strategy of the WMAP satellite is shown in Fig.\,\ref{fig:WMAPscan-1}.
The blue arrows show the lines of sight of the two telescopes.
These two line of sight directions are denoted by $\hat{R}_{1}$ and
$\hat{R}_{2}$.
In the ecliptic coordinate system these direction vectors 
can be obtained as a function of time as

\begin{eqnarray}
\hat{R}_{1,2} & = & \left[\begin{array}{ccc}
\cos\omega_{1}t & \sin\omega_{1}t & 0\\
-\sin\omega_{1}t & \cos\omega_{1}t & 0\\
0 & 0 & 1
\end{array}\right]\left[\begin{array}{ccc}
1 & 0 & 0\\
0 & \cos\omega_{2}t & \sin\omega_{2}t\\
0 & -\sin\omega_{2}t & \cos\omega_{2}t
\end{array}\right]\left[\begin{array}{ccc}
\cos\alpha & 0 & \sin\alpha\\
0 & 1 & 0\\
-\sin\alpha & 0 & \cos\alpha
\end{array}\right] 
 \left[\begin{array}{c}
\cos\beta\\
\pm\sin\beta\cos\omega_{3}t\\
\pm\sin\beta\sin\omega_{3}t
\end{array}\right]\,,\label{eq:line_of_sight1}
\end{eqnarray}

\noindent where the last matrix should take the positive sign (`+') for $\hat{R}_{1}$ and negative sign (`-') for $\hat{R}_{2}$. 
Here $\alpha, \beta, \omega_{1}, \omega_{2}$ and $\omega_{3}$
are as shown in the schematic diagram given in Fig.\,\ref{fig:WMAPscan-1}.
The values of these angles and the rotation rates are 
$\alpha=22.5^{\circ}$, 
$\beta=70.5^{\circ}$, 
$\omega_{3}=2.915398\,$ rad/min, 
$\omega_{2}=0.104719755\,$ rad/min, 
$\omega_{1}=0.000011954\,$ rad/min.

The perpendicular direction to both $\hat{R}_{1}$ and $\hat{R}_{2}$ vector is given by their cross product as  

\begin{equation}
\hat{Z}=-\hat{R}_{1}\times\hat{R}_{2}\;.
\end{equation}

\noindent The negative signature is to be consistent with the schematic diagram  
presented in Fig.\,\ref{fig:WMAPscan-1}. At this point it should be noted that $\hat{Z}$ and $-\hat{Z}$ are the scan axies for the beams in $\hat{R}_{1}$ and $\hat{R}_{2}$ directions respectively. 

For measuring the beam orientation we need the perpendicular planes say $\zeta_1$ and $\zeta_2$ to the vectors $\hat{R}_{1}$ and $\hat{R}_{2}$ respectively and two perpendicular directions lying on each of those planes. As $\hat{Z}$ 
is perpendicular to both $\hat{R}_{1}$ and $\hat{R}_{2}$, it is a common direction in both the planes  $\zeta_1$ and $\zeta_2$. 

If we consider  $\hat{I}_{1}$ and $\hat{I}_{2}$ to be the directions perpendicular to $\hat{Z}$  on the plane $\zeta_1$ and $\zeta_2$ respectively then they will be given by

\begin{equation}
\hat{I}_{i}=\hat{R}_{i}\times\hat{Z}\;\quad \ldots\;\forall i\in\{1,2\}\;. 
\end{equation}

\noindent  If we consider that the X-axes of the beam sensitivity functions are aligned 
at an angle $\varphi_{1}$ and $\varphi_{2}$ with the directions $\hat{I}_{1}$and $\hat{I}_{2}$ respectively, then the directions of  the  X-axes of the beams in our coordinate system will be given by $\hat{N}_{1}$ and $\hat{N}_{2}$ respectively, where 

\begin{equation}
\hat{N}_{i}=\cos\varphi_{i}\hat{I}_{i}+\sin\varphi_{i}\hat{Z}\;\quad \ldots\;\forall i\in\{1,2\}\;.\label{eq:semimajor1}
\end{equation}

\noindent In general the beam sensitivity functions are given in the scan coordinate system, where scan direction is taken as the beam X-axes and cross scan direction as beam Y-axes. In such a case $\varphi_1=90^o$ and $\varphi_2=-90^o$ respectively. 

At this point we should note that in section~\ref{section2}, the coordinate system is chosen in such a way that the equation for dipole becomes $T_{0}Y_{1}^{0}(\gamma)$, i.e. the dipole is oriented along the $z-axis$ of the coordinate system. However, the scan strategy
discussed here is in the ecliptic coordinate system. Therefore, for the beam convolution, we need to transform the vectors $\hat{R}_{1}$, $\hat{R}_{2}$, $\hat{N}_1$ and $\hat{N}_2$ in the dipole coordinate system (the coordinate system where
the $z-axis$ is oriented along the dipole) which can be done by multiplying with the rotation matrix

\begin{equation}
\tilde{W}=\left[\begin{array}{ccc}
-0.9703& \;\;0.1394&  -0.1974\\
-0.1036& -0.9779&  -0.1815\\
-0.2184& -0.1557& \;\;0.9634\\
\end{array}\right]\;.
\end{equation}

Using Eq.\ref{eq:line_of_sight1} we can find out the directions of the beams i.e. $\hat{R}_{1}$ and
$\hat{R}_{2}$ that give us the $\theta_1$ and $\theta_1$ for the beams. Using Eq.\ref{eq:semimajor1} we can calculate the physical orientation of the beams i.e $\rho_1$ and $\rho_2$ which are the angles between the $\hat{N}_1$ and $\hat{N}_2$ with the local longitudinal direction. 
Using these we can calculate the $\cos\theta_{1}$, $\cos\theta_{2}$, $\sin\theta_{1}$$\cos\rho_{1}$,  $\sin\theta_{2}$$\cos\rho_{2}$, $\sin\theta_{1}$$\sin\rho_{1}$ and $\sin\theta_{2}$$\sin\rho_{2}$ respectively. In these
expressions the subscript $1$ and $2$ represents the two beam. 
Using Eq.\,\ref{eq:temp} we can calculate three independent TOD for each beam independently. 
Multiplying three independent TODs with the beam
harmonic coefficients $b_{10}(\hat{z})$, $b_{r}(\hat{z})$ and $b_{i}(\hat{z})$, and 
then using some map-making technique we can thus calculate three independent  
maps. These maps can be summed up to get a final map, which can then be analyzed to know 
the total amount of power leakage.

\subsection{WMAP Beam functions}

The WMAP satellite scans the sky temperature in 5 different frequency bands, named as $K$, $Ka$, $Q$, $V$ and $W$. Amongst them $Q$ and $V$ band have two detectors each and $W$ band has four detectors. Each of these detectors has one pair of beams. The beam maps are made from the in-flight observations of the Jupiter. The beam sensitivity functions $b(x_b,y_b)$ are given in a $600\times 600$ grid
with the center at the optic axis of the differential assembly (DA), where $x_b$ and $y_b$ are the axes in the beam map. The observations of the antenna temperature are given in ${\rm mK}$ and assigned in $2.4\times2.4\, {\rm arcmin}^2$ beans. 

\begin{figure}
\includegraphics[width=0.47\textwidth]{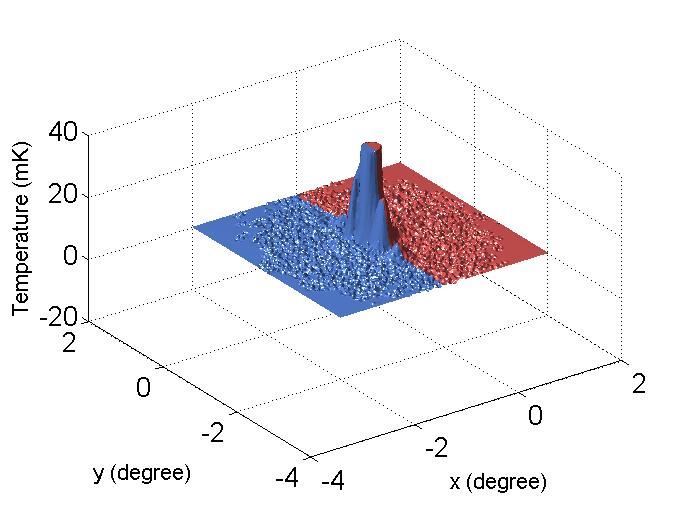}
\includegraphics[width=0.47\textwidth]{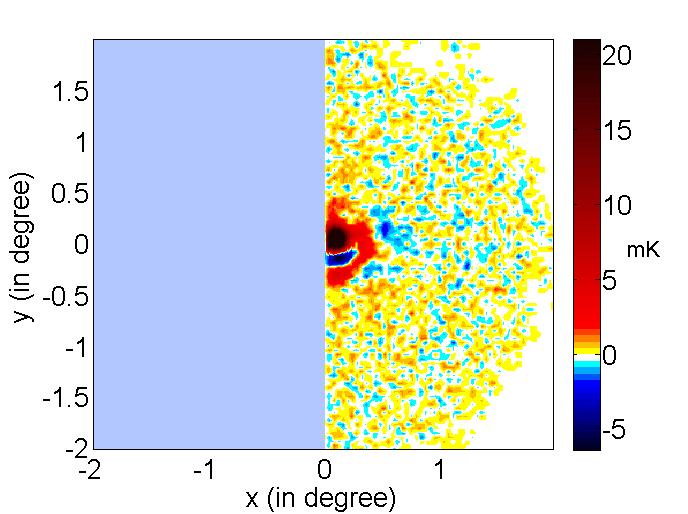}
\centering
\includegraphics[width=0.47\textwidth]{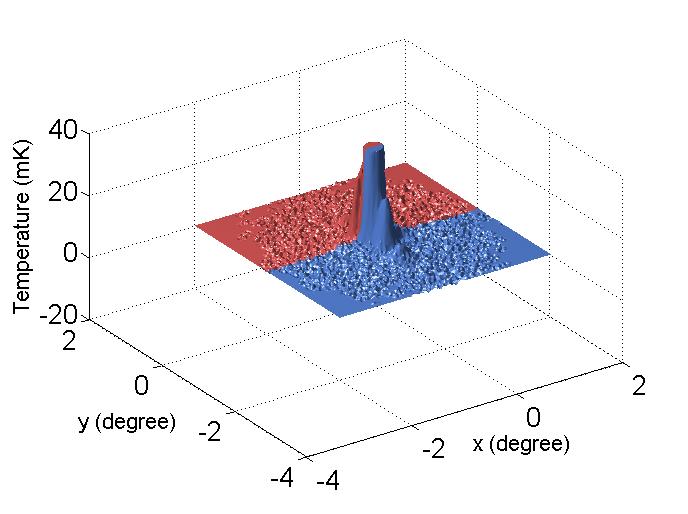}
\includegraphics[width=0.47\textwidth]{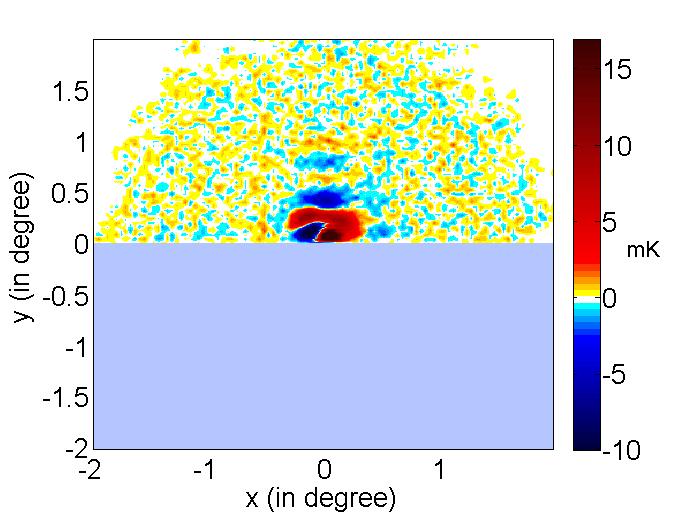}

\caption{\label{fig:W1A}The figures in the left show the enlarged plot of the A side beam of W3 band. 
We choose the maximum sensitivity point as the center of the beam. The central peak of the beam 
is truncated after $30$mK in the plot. We plot one part of the beam in red and other in blue (top division along 
X axis, bottom division along Y axis). Visibly it is clear that 
two sides of the beam are not symmetric ( mirror reflection of one another ).  In the right figures 
we have subtracted one half of the beam from the another half (red part - blue part) to show the asymmetry 
in the beam. 
}
\end{figure}

None of these beams are reflectional symmetric. 
In Fig.\,\ref{fig:W1A}  we show the zoomed up view of the W3A side beam map, where we truncate the central peak after a sensitivity of $30\,{\rm mK}$. 
The figure shows that apart from the central peak there are many other small structures in 
the beam. There are small shoulders and an annular noisy region.  
These lead to a nonzero value of the parameters $b_{r}(\hat{z})$ and $b_{i}(\hat{z})$. In left column of Fig.\,\ref{fig:W1A}
we plot one-half of the beam in red and the other in blue, divided along X and Y axis of the beam. 
We choose the center of the peak sensitivity pixel of the beam as the center of the coordinate system. The plots show that there are some clear asymmetry between the two halves of the beam.  To show the asymmetry explicitly we subtract the blue half from the red half and plot it in the right panel of the figure. 

Given a beam sensitivity function, the harmonic coefficients of the beam can be calculated by integrating it with the corresponding conjugate spherical harmonics. This gives

\begin{eqnarray}
b_{10}(\hat{z})&=&\int b(\hat{z},\gamma)Y_{10}^{*}(\gamma)d\gamma\;, \\
b_{r}(\hat{z})&=&\int b(\hat{z},\gamma)\frac{1}{2}\left(Y_{1,1}(\gamma)-Y_{1,-1}(\gamma)\right)d\gamma\;, \\
b_{i}(\hat{z})&=&\int b(\hat{z},\gamma)\frac{i}{2}\left(Y_{1,1}(\gamma)+Y_{1,-1}(\gamma)\right)d\gamma\;. 
\end{eqnarray}

\begin{table}

\begin{center}

\setlength{\tabcolsep}{8pt}
\renewcommand{\arraystretch}{1.4}
\begin{tabular}{|ScScScScScScScScScScSc|}
\hline
$\,$  & $b_{i}^{A(P)}$  & $b_{r}^{A(P)}$  & $b_{i}^{B(P)}$  & $b_{r}^{B(P)}$  & $T_{q}^{(P)}$ & $b_{i}^{A(W)}$  & $b_{r}^{A(W)}$  & $b_{i}^{B(W)}$  & $b_{r}^{B(W)}$  & $T_{q}^{(W)}$\tabularnewline
$K_{1}$ & $2.56$ & $\llap{$-$}1.44$ & $2.66$ & $\llap{$-$}0.24$ & $2.52$ & $2.80$ & $1.60$ & $2.01$ & $1.24$ & $\llap{$-$}2.17$\tabularnewline
$Ka_{1}$ & $\llap{$-$}1.36$ & $\llap{$-$}1.78$ & $1.92$ & $1.32$ & $0.69$ & $\llap{$-$}0.19$ & $0.83$ & $1.20$ & $2.10$ & $\llap{$-$}2.24$\tabularnewline
$Q_{1}$ & $2.73$ & $\llap{$-$}2.45$ & $2.25$ & $2.23$ & $0.34$ & $0.26$ & $\llap{$-$}0.66$ & $0.50$ & $1.85$ & $\llap{$-$}0.99$\tabularnewline
$Q_{2}$ & $1.31$ & $2.58$ & $1.03$ & $\llap{$-$}3.27$ & $1.07$ & $\llap{$-$}0.88$ & $1.99$ & $2.17$ & $1.12$ & $\llap{$-$}2.38$\tabularnewline
$V_{1}$ & $0.97$ & $\llap{$-$}0.57$ & $1.30$ & $0.50$ & $0.12$ & $\llap{$-$}0.02$ & $0.61$ & $\llap{$-$}0.56$ & $\llap{$-$}0.96$ & $\llap{$-$}1.20$\tabularnewline
$V_{2}$ & $0.84$ & $2.03$ & $0.53$ & $\llap{$-$}1.94$ & $\llap{$-$}0.14$ & $\llap{$-$}0.79$ & $1.52$ & $\llap{$-$}0.48$ & $0.41$ & $\llap{$-$}1.47$\tabularnewline
$W_{1}$ & $\llap{$-$}0.00$ & $\llap{$-$}0.89$ & $0.71$ & $0.08$ & $1.30$ & $\llap{$-$}1.01$ & $1.17$ & $\llap{$-$}1.29$ & $1.15$ & $\llap{$-$}1.98$\tabularnewline
$W_{2}$ & $1.12$ & $0.41$ & $2.45$ & $\llap{$-$}0.73$ & $0.49$ & $\llap{$-$}0.86$ & $1.34$ & $0.13$ & $0.73$ & $\llap{$-$}1.58$\tabularnewline
$W_{3}$ & $0.06$ & $0.39$ & $1.74$ & $0.81$ & $\llap{$-$}1.85$ & $\llap{$-$}0.48$ & $0.50$ & $\llap{$-$}0.33$ & $0.79$ & $\llap{$-$}0.98$\tabularnewline
$W_{4}$ & $\llap{$-$}0.03$ & $0.46$ & $0.017$ & $9.27$ & $\llap{$-$}0.87$ & $\llap{$-$}1.89$ & $0.57$ & $\llap{$-$}0.04$ & $0.66$ & $\llap{$-$}0.94$\tabularnewline
 \hline
\end{tabular}

\end{center}

\caption{\label{tab:The-coefficients}The dipole coefficients,  $b_{i}$ and $b_{r}$ (in $10^{-4}$ order)
of the beam spherical harmonics for different WMAP beams 
estimated from the publicly available beam maps \cite{beam}. The quadrupole temperatures ($\sqrt{3C_2/\pi}$ in $\mu$K unit)
due to dipole power leakage are calculated. The values with superscript `P' are calculated considering 
the center of the peak sensitivity pixel as LOS direction and with superscript `W' are calculated with WMAP-9 LOS direction}
\end{table}

\noindent Here $\gamma$ is the direction vector. As the beam map is provided in a $2.4\,{\rm arcmin}$ grid, we need to convert $\gamma$ in the Cartesian system i.e. ($x_b$, $y_b$) while doing the integration. $\hat{z}$ is the direction of the line of sight. We can use the center of the peak sensitivity pixel as the line of sight of the beam and calculate the power leakage. However, even a slight shift the line of sight direction may significantly change the values of $b_{r}(z)$ and $b_{i}(z)$ and hence the power leakage. 
The WMAP-9 team calculate the lines of sight directions using Jupiter map 
which is slightly different from the center of the peak sensitivity pixel. For different bands the WMAP-9 line of sight direction differs from the center of the peak sensitivity direction differs by $0'-2'$. We do the analysis for both the line of sight. The amount of power leakage for both the cases are shown in table (\ref{tab:The-coefficients}).

At this point it can be noted that if the centroid of the beam i.e. $\bar{x}_{b} = \frac{\iint b(x_{b},y_{b})x_{b}dx_{b}dy_{b}}{\iint b(x_{b},y_{b})dx_{b}dy_{b}}$ and $\bar{y}_{b} = \frac{\iint b(x_{b},y_{b})y_{b}dx_{b}dy_{b}}{\iint b(x_{b},y_{b})dx_{b}dy_{b}}$ is taken as the pointing direction of the beam then the amount of power leakage becomes almost negligible for the WMAP beams. 
Though considering a separate line of sight can introduce other complex features in the observed sky-map as shown in \cite{Joshi2012,Das2014a,Das2015}, which can only be understood through proper simulations.

\subsection{WMAP Map making}

WMAP uses  differential scan pattern i.e. it measures the temperature difference of the sky between two points separated by $\sim141\degree$. This time order data (TOD) is then used  to construct the map using proper map making techniques~\cite{Hamilton-2003}. WMAP mapmaking technique can be briefly described as follows. If the actual temperature of the pixelated sky at a given direction is given by ${\bf\tilde{T}}$ vector then 

\begin{equation}
{\bf\tilde{d}}={\bf A}{\bf\tilde{T}}+{\bf\tilde{T}_{n}}\;.
\label{eq:Pointing}
\end{equation}

\noindent where, ${\bf\tilde{T}}$ is the map vector of length $N_p$ and ${\bf\tilde{d}}$ is the TOD vector of length $N_t$. ${\bf A}$ is a matrix of order $N_{t}\times N_{p}$ and is known as the pointing matrix. Any row of ${\bf A}$ matrix consists of only two nonzero elements, one with $+1$ and the other with $-1$, corresponding to two pixels in the map that got scanned at that particular time in the sky.  Here $N_{t}$ is the number of elements in the TOD and $N_{p}$ is
the number of pixels in the map.

For our analysis we consider ${\bf\tilde{T}_{n}}=0$. The estimated sky-map, ${\bf\hat{\tilde{T}}}$ can be calculated from the TOD by 
taking the pseudo inverse of the matrix ${\rm A}$, i.e. 

\begin{equation}
{\bf \hat{\tilde{T}}}=[{\bf A}^{T}{\bf A}]^{-1}{\bf A}^{T}{\bf\tilde{d}}\;.
\end{equation}

\noindent It is not time efficient to invert the matrix $[{\bf A}^{T}{\bf A}]^{-1}$ using brute force method. Therefore, we use Jacobi iteration 
to solve the equation ${\bf\tilde{d}}={\bf A}{\bf\hat{\tilde{T}}}$. 
The equation used in the iteration method \cite{Hamilton-2003} is

\begin{equation}
{\bf\hat{\tilde{T}}_{k+1}}={\bf\hat{\tilde{T}}_{k}}+[\rm diag({\bf A}^{T}A)]^{-1}{\bf A}^{T}({\bf\tilde{d}}-{\bf A}{\bf\hat{\tilde{T}}_{k}})\label{eq:iteration}\;,
\end{equation}

\noindent where the suffix $k$ stands for the $k^{th}$ iteration step. For initializing the 
iteration we take ${\bf\hat{\tilde{T}}_{0}}$ as a dipole map similar to 
the dipole used in the exercise. Here `tilde' is used to denote vectors (1D matrices) and we write all the matrices in bold.

\section{Simulation and Results from WMAP scan pattern}
\label{sec:WMAP}

For analyzing the amount of power leakage, we use a dipole map, similar to the CMB dipole, which in the galactic coordinate is oriented along $(l,b)=(264.31^{o},48.05^{o})$ with an amplitude of $3.358\,$mK. The WMAP satellite scans the full sky once in 6 months. However, we consider  a full year scan to guarantee all the symmetries. Healpix-2.15 \cite{healpix} is used for carrying out
all the analysis. We use $N_{side}=256$.

\begin{figure}
\centering

\subfloat[\label{fig:WMAP-simulation}  The figures in the left panel show three independent 
maps from $\cos\theta$, $\sin\theta\cos\rho$ and $\sin\theta\sin\rho$ components.
The scanned dipole map is a linear combination of these three maps.
The figures in the middle panel show the shapes of the quadrupole produced
by the dipole power leakage for all 10 different WMAP bands (considering peak sensitivity direction of 
the beams as LOS). Right top figure shows the quadrupole 
observed from the WMAP 9 year ILC map. The colorbar show the temperature 
in the ${\rm mK}$ unit. The plots show that the shape of the quadrupoles generated 
by the dipole power leakage are similar to the quadrupole observed by WMAP satellite. All these maps are shown in the Ecliptic 
coordinate system. The bar diagram at the right bottom shows the amount of temperature leaked to quadrupole 
by different WMAP bands.]{\includegraphics[width=0.98\textwidth]{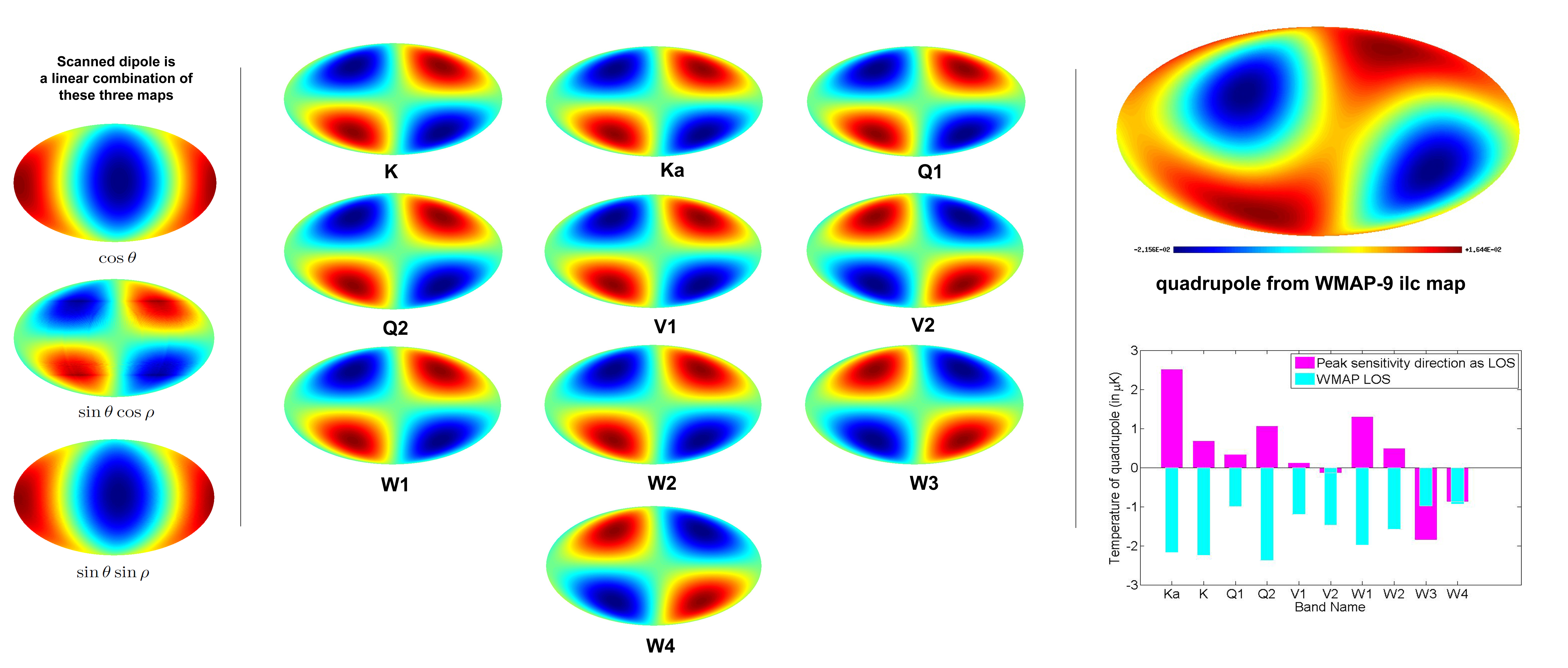}

}

\subfloat[\label{fig:quadrupolealignment}
$n$ times of the temperature map obtained by dipole leakage for W3 band is added to the $1000$
random maps generated by Healpix and their quadrupole orientations are analyzed. 
The X-axis shows the $|\cos(\theta_{n_{1}n_{2}})|$, where $n_{1}$ is the quadrupole direction from the simulated maps
and $n_{2}$ is the direction of the quadrupole from the WMAP-9 year ILC map. Y-axis
shows the number of maps in each bin where bin size is $0.1$.]{\includegraphics[width=0.98\textwidth]{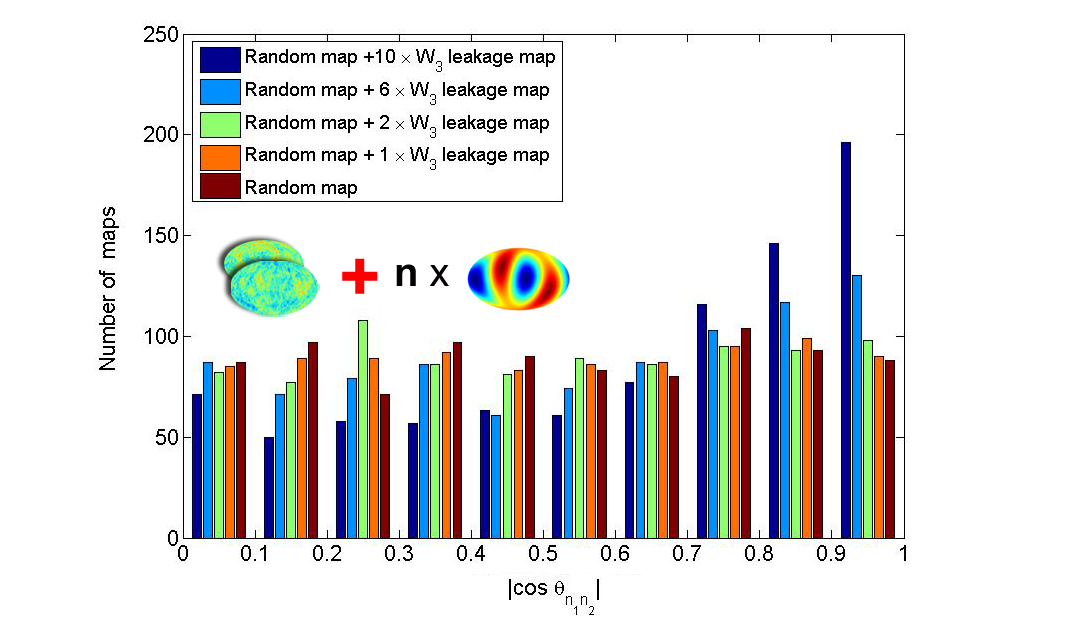}

}

\caption{Analysis for the WMAP scan strategy}

\end{figure}

We use the method described in section (\ref{section2}) and section (\ref{sec3}) for generating the TOD and then use the map making technique to reconstruct the final map. Our analysis shows that the temperature leaked from dipole into quadrupole varies for different beam maps as $b_r$ and $b_i$ are different for different beams. In Fig.\,\ref{fig:WMAP-simulation} we show the shape of the quadrupole generated due to the the power leakage from dipole for all the bands of the WMAP satellite. We also show the shape of the quadrupole component 
from the WMAP ILC map in the right column. It is interesting that the shape of the quadrupole arising from the dipole power leakage is 
oriented in the same way as that of the measured WMAP quadrupole.  The direction of quadrupole can be calculated by maximizing $\sum_{m}m^{2}|a_{2m}^{2}(\hat{n}_1)|$ with respect to the direction $\hat{n}_1$ \cite{Ade2013,Tegmark2003,deOliveiraCosta2003}
In the ecliptic coordinate system the WMAP quadrupole is oriented along $(342.95\degree,82.82\degree)$ whereas the the quadrupole orientation from the dipole power leakage for 'K' band is along $(351.74\degree,89.70\degree)$, which are strikingly very close to each other. As most the quadrupole for the dipole leakage is coming from the $\sin\theta\cos\rho$ component of the map, therefore, the orientation of the quadrupole for all the other beams are also almost same. 
Although for some of the beams the phases are opposite to the WMAP ILC map quadrupole, but in all the cases the observed quadrupoles definitely get modified significantly by the power leakage observed here. 

The values of the beam spherical harmonic coefficients i.e. $b_{r}$, $b_{i}$ along 
with the amount of temperature leakage $\left(\sqrt{3C_2/\pi}\right)$, for different bands are listed in table \ref{tab:The-coefficients}.  We show the amount of the temperature leakage to the quadrupole for both peak sensitivity direction and the WMAP simulated line of sight. We can see that amount of the power leakage differs significantly for a slight change in the line of sight. In some of the cases the amount of the power leakage is significant in comparison to the observed CMB quadrupole which is $\sim 12.27\mu{\rm K}$ (WMAP-9). For $K$ band (with Peak sensitivity direction as line of sight direction) the amount of temperature modification after scanning can be $\sim14.7\mu {\rm K}$ which leads to a power change of $\sim 65.5 \mu {\rm K}^2$.

To check if this  quadrupole generated due to the power leakage from dipole can affect 
the orientation of the quadrupole in a random map, we generate 1000 random realizations 
from $C_l$ obtained from WMAP-9 year data. We add $n$ times of the dipole power leakage map from W3 band
of the WMAP-satellite with the random realizations and then calculate the quadrupole orientations $\hat{n}_1$ in those maps. 
The cosine of these directions with the quadrupole direction of WMAP
ILC map ( $\hat{n}_2$)  is calculated and then the maps are distributed in 10 bins each of size 0.1 according to the cosine values. 
The significance of taking $n$-times of the W3 leakage map is that a very small amount of change in the pointing vector can change 
the amount of power leakage. Therefore, by doing the analysis for different $n$ we can explicitly say how much temperature leakage can statistically explains the quadrupole orientation anomaly. 
Also, the effect of any pointing offset will be encompassed by this factor. 

In Fig.\,\ref{fig:quadrupolealignment} we show the results of our analysis.
For $n=1$, there is no statistically significant direction anomaly. Therefore, we can ignore the possibility of direction anomaly only due the dipole power leakage. 
However, if the quadrupole is even $6$ times the quadrupole generated by W3 band, i.e. if this external quadrupole is almost of similar value 
of WMAP quadrupole ($\sim12.0 \mu {\rm K}$), then there is a very high probability that the final quadrupole will be aligned towards the quadrupole direction WMAP ILC map . 

At this point it should also be noted that  for WMAP scan pattern 
the power leakage is very much restricted to the quadrupole. The amount of the temperature goes to the 
octupole or the higher multipoles are very less (octupole temperature amplitude is about 100 times lesser than quadrupole) in comparison to the quadrupole component.

\begin{figure}
\centering
\subfloat[\label{fig:Planck-simulation} $T_0\cos\theta$, $T_0\sin\theta\cos\rho$
and $T_0\sin\theta\sin\rho$ maps for Planck scan strategy in $\mu$K. All the maps are shown in the ecliptic coordinate system.]{\includegraphics[width=0.32\textwidth]{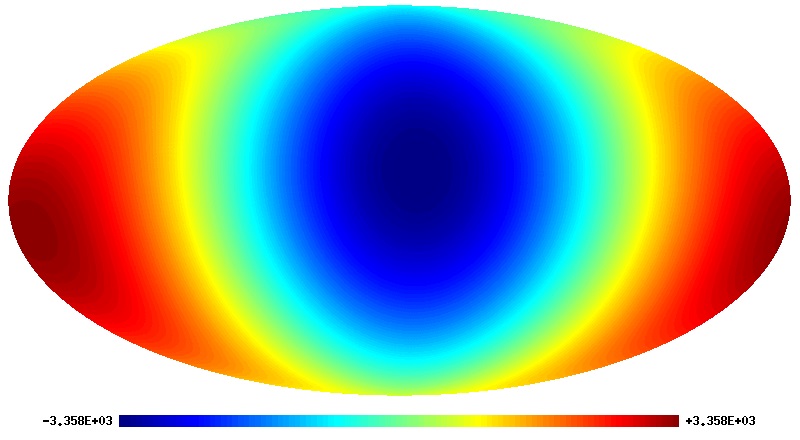}\includegraphics[width=0.32\textwidth]{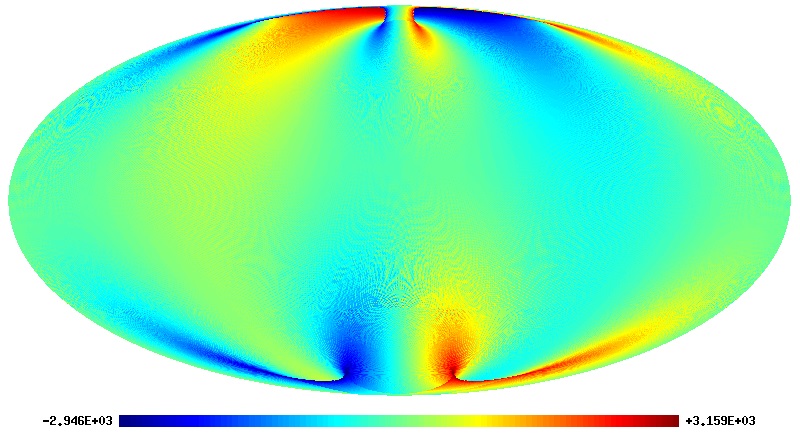}\includegraphics[width=0.32\textwidth]{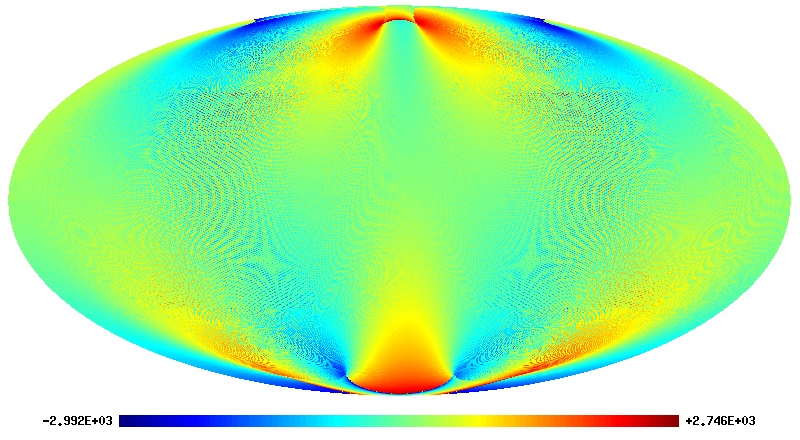}
}

\subfloat[\label{fig:Planck_break}Amount of power leakage to different multipoles considering the leakage to quadrupole to be unity.
Unlike WMAP here the power leakage is not restricted to the quadrupole and hence if the beam asymmetry parameters are big
the dipole power leakage will modify all the low multipoles. We also show the shape of the quadrupole and the octupole
part from the $T_0\sin\theta\sin\rho$ and $T_0\sin\theta\cos\rho$ components.]{
\includegraphics[width=0.54\textwidth]{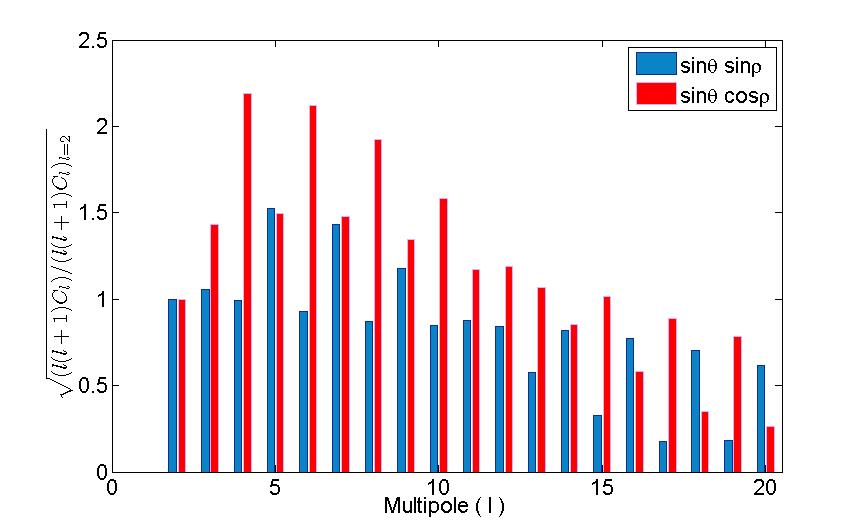}
\includegraphics[width=0.420\textwidth,trim = 105 235 105 235, clip]{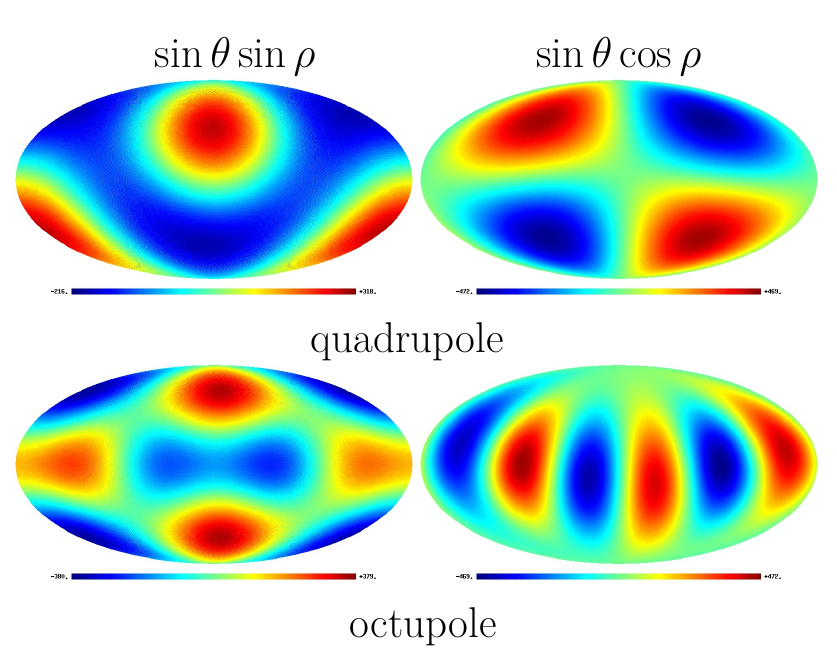}
}

\caption{Analysis for the Planck scan strategy}

\end{figure}

\section{Simulations for Planck scan strategy and results}
\label{sec:Planck}

Planck satellite follows a different scan strategy than that of the
WMAP\cite{BlueBook2005}. It has only one beam and instead of the differential
measurement it directly measures the temperature of the sky. The Planck
satellite beam is approximately $85\degree$ off symmetric axis and the
precession angle for the satellite is around $7.5\degree$. The precession
rate is taken as one revolution per six month and the spin rate as
$1{\rm rpm}$. A similar set of steps as that of WMAP can also be followed for
analyzing the Planck scan strategy. As Planck has only one beam, its orientation can be found out by the direction $R_1$ 
in Eq.\ref{eq:line_of_sight1}. Planck time ordered data consists of measured temperature in a particular 
sky direction instead of the temperature difference as in case of WMAP and the Planck pointing matrix in Eq.\ref{eq:Pointing} consists of only one $+1$
component in each of the row.

Using the similar techniques as described in section (\ref{section2}) and section (\ref{sec3}), three independent maps of  $T_0 \cos\theta$, $T_0 \sin\theta\cos\rho$ and $T_0 \sin\theta\sin\rho$ can be calculated for Planck scan strategy. These maps are shown
 in Fig.\,\ref{fig:Planck-simulation}. The full map after scanning the dipole is given by summing the three maps with the proper weight as shown in Eq.\ref{eq:temp}. 
These  maps show that for Planck scan strategy the power leakage is not only restricted to the quadrupole. 
The power gets transferred to all other higher low multipoles and affects the low-multipole
temperature measurement. The amount of power transfer  depends on the coefficients $b_{i}$ and $b_{r}$. 
In left panel of  Fig.\,$\ref{fig:Planck_break}$ we plot the amount of temperature leakage to octupole and higher multipole 
in parts of quadrupole temperature amplitude i.e. $\sqrt{\frac{l(l+1)C_l}{(l(l+1)C_l)_{l=2}}}$ for the  $T_0 \sin\theta\cos\rho$ and $T_0 \sin\theta\sin\rho$ maps.
In the right panel of Fig.\,$\ref{fig:Planck_break}$ we show the shape of the quadrupole and octupole in the scanned maps. 
The quadrupole temperature $(\sqrt{3C_2/\pi})$  for $T_0 \sin\theta\cos\rho$ and $T_0 \sin\theta\sin\rho$ maps are $ b_r \times  360.49\mu{\rm K}$ and $ b_i \times 203.94\mu{\rm K}$ respectively. 
Therefore, if the effect of the beam during map-making is not properly taken in account then these features can contaminate the resultant map and can induce isotropy violation signal at low multipoles.
If the values of $b_{r}$ and $b_{i}$ are of the order of $10^{-4}$ then the amount of the temperature leakage to quadrupole and other high multipoles will be of $O(10^{-1})\mu{\rm K}$, which can be neglected. However, if  $b_{r}$  and $b_{i}$  are higher then the contamination can affect the Planck results significantly in view of the accuracy promised by the Planck satellite.

\section{Conclusion}

We develop an analytical formalism to simulate WMAP and Planck the scan strategy and use it 
to estimate the leakage of power from dipole. We show that the power leakage
only depends on the two spherical harmonic coefficients of the satellite
beam ($b_{i}$ and $b_{r}$) and if the beam is designed
in such a way that these two parameters of the beam are small enough then power 
leakage will be negligible. For WMAP, the amount of the power leakage
is found to be small but not insignificant compared to the low value of quadrupole measured. 
The quadrupole shape generated by the dipole power leakage is similar to the observed quadrupole in CMB sky. 
The leakage of power depends on the scan pattern. However, our analysis show that the quadrupole direction anomaly 
is not possible to explain using only this particular effect. For WMAP scan pattern the power leakage 
is mainly restricted to the quadrupole. However, for Planck scan pattern the power gets leaked to 
quadrupole and octupole and other higher low multipoles. Though for Planck scan strategy the amount of the power leakage is much 
smaller in comparison to the WMAP. 
This particular analysis will definitely help beam design for the future high precession CMB experiments.

\acknowledgments{
We have used the HPC facility at IUCAA for the required computation. SD acknowledge Council for Science and Industrial Research (CSIR), India, for the financial support as Senior Research Fellows.}

\bibliographystyle{JHEP}
\bibliography{references}

\end{document}